# 19.3 GHz Acoustic Filter with High Close-in Rejection in Tri-layer Thin-Film Lithium Niobate

Omar Barrera, Sinwoo Cho, Jack Kramer, Vakhtang Chulukhadze, Tzu-Hsuan Hsu, and Ruochen Lu

***Abstract*— Acoustic filters are preferred front-end solutions at sub-6 GHz due to their superior frequency selectivity compared to electromagnetic (EM) counterparts. With the ongoing development of 5G and the evolution toward 6G, there is a growing need to extend acoustic filter technologies into frequency range 3 (FR3), which spans 7 to 24 GHz to accommodate emerging high-frequency bands. However, scaling acoustic filters beyond 10 GHz presents significant challenges, as conventional platforms suffer from increased insertion loss (IL) and degraded out-of-band (OoB) rejection at higher frequencies. Recent innovations have led to the emergence of periodically poled piezoelectric lithium niobate (P3F LN) laterally excited bulk acoustic resonators (XBARs), offering low-loss and high electromechanical coupling performance above 10 GHz. This work presents the first tri-layer P3F LN filter operating at 19.3 GHz, achieving a low IL of 2.2 dB, a 3-dB fractional bandwidth (FBW) of 8.5%, and an impressive 49 dB close-in rejection. These results demonstrate strong potential for integration into FR3 diplexers.**

Table I State-of-the-Art Acoustic Filters in FR3

| Reference | $f_c$ (GHz) | *IL* (dB) | *FBW* (%) | Close-in Rejection (dB) |
|---|---|---|---|---|
| [14] | 8.7 | 0.7 | 6.08 | 8.2 / 12.25 |
| [15] | 9.2 | 0.76 | 5.1 | 3.8 / 4.2 |
| [16] | 17.4 | 3.25 | 3.4 | 20.7 / 18.05 |
| [13] | 19.1 | 8.1 | 3.1 | 13.5 / 13.2 |
| [19] | 23.5 | 2.38 | 18.2 | 16.4 / 21.27 |
| [22] | 23.8 | 1.52 | 19.4 | 17.23 / 22.66 |
| **This work** | **19.5** | **2.2** | **9.6** | **49.9 / 14** |

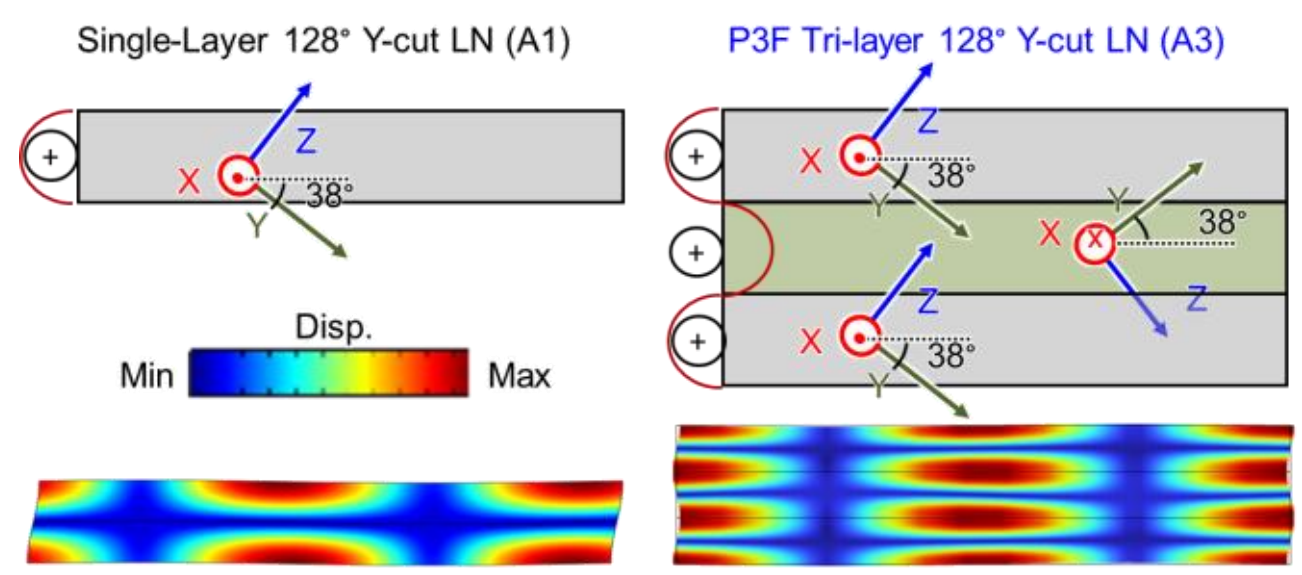

Fig. 1. Schematic comparison between single layer and P3F tri-layer 128-Y cut LN (this work), operating as A1 and A3 XBARs.

## I. Introduction

RADIO FREQUENCY (RF) spectrum is becoming increasingly crowded as more wireless applications are envisioned for mobile devices. This trend creates a growing need for pre-selection of electromagnetic (EM) signals, leading to the pervasiveness of RF front ends [1]. Among front-end components, RF filters provide the pre-filtering of EM signals for both transmitters and receivers, thus low-loss and high-rejection filters within compact sizes are highly sought-after [2], [3]. Conventionally, such filters are achieved with piezoelectric devices [4], [5], e.g., surface acoustic wave (SAW) filters based on lithium tantalate (LT) and lithium niobate (LN) [6], or bulk acoustic wave (BAW) filters [7] based on aluminum nitride (AlN) and scandium aluminum nitride (ScAlN) [8]–[10]. These technologies have been widely adopted and proven effective in sub-6 GHz bands. With wireless standards now moving toward higher frequencies, more recently, frequency range 3 (FR3, 7.125 GHz to 24.25 GHz) have been identified for 6G [11], [12] wireless communication, causing new challenges for filters, as conventional filter performance degrades. Consequently, new platforms are needed.

Recently, acoustic filters beyond 8 GHz have been demonstrated in LN [13] and ScAlN [14]–[16]. In the case of LN, filters using lateral-field excited bulk acoustic wave resonators (XBARs) [17], [18], showing low insertion loss (IL) and wide fractional bandwidth (FBW) [19]. These advancements are enabled by first-order antisymmetric (A1) mode resonators (Fig. 1) with high quality factor (Q) and coupling ($k^2$). More recently, periodically poled piezoelectric films (P3F) in LN [20], where multiple layers of LN are transferred successively with opposite polarizations (Fig. 1), have enabled the excitation of higher-order Lamb modes in thicker films [21], resulting in higher quality factors (Q) and further improvements in insertion loss (IL) and fractional bandwidth (FBW) [22]. Nevertheless, such prototypes beyond 10 GHz have been focused on low IL so far, without emphasis on out-of-band rejection (OoB), especially close-in rejections (Table I), crucial for applications like diplexers. The lack of high-rejection filters is mostly caused by the fabrication challenge of precisely setting the resonances of multiple resonators, required for introducing transmission zeros and poles toward filters of better rejection [23]. This limitation highlights the need for alternative approaches to achieve high-rejection filter designs.

In this work, we achieve high close-in rejection in LN filters by leveraging the multi-mode behavior of P3F stacks, which occurs naturally from layer mismatches introduced during processing. The adjacent modes introduce additional transmission zeroes next to the passband, significantly enhancing OoB rejection in a moderate 5th order filter. More specifically, we present the first trilayer P3F LN filter at 19.3 GHz, showing a low IL of 2.2 dB, a 3-dB FBW of 8.5%, and a remarkable 49.9 dB close-in rejection, within a footprint of 0.95 mm². Upon development, such results are promising for FR3 applications.

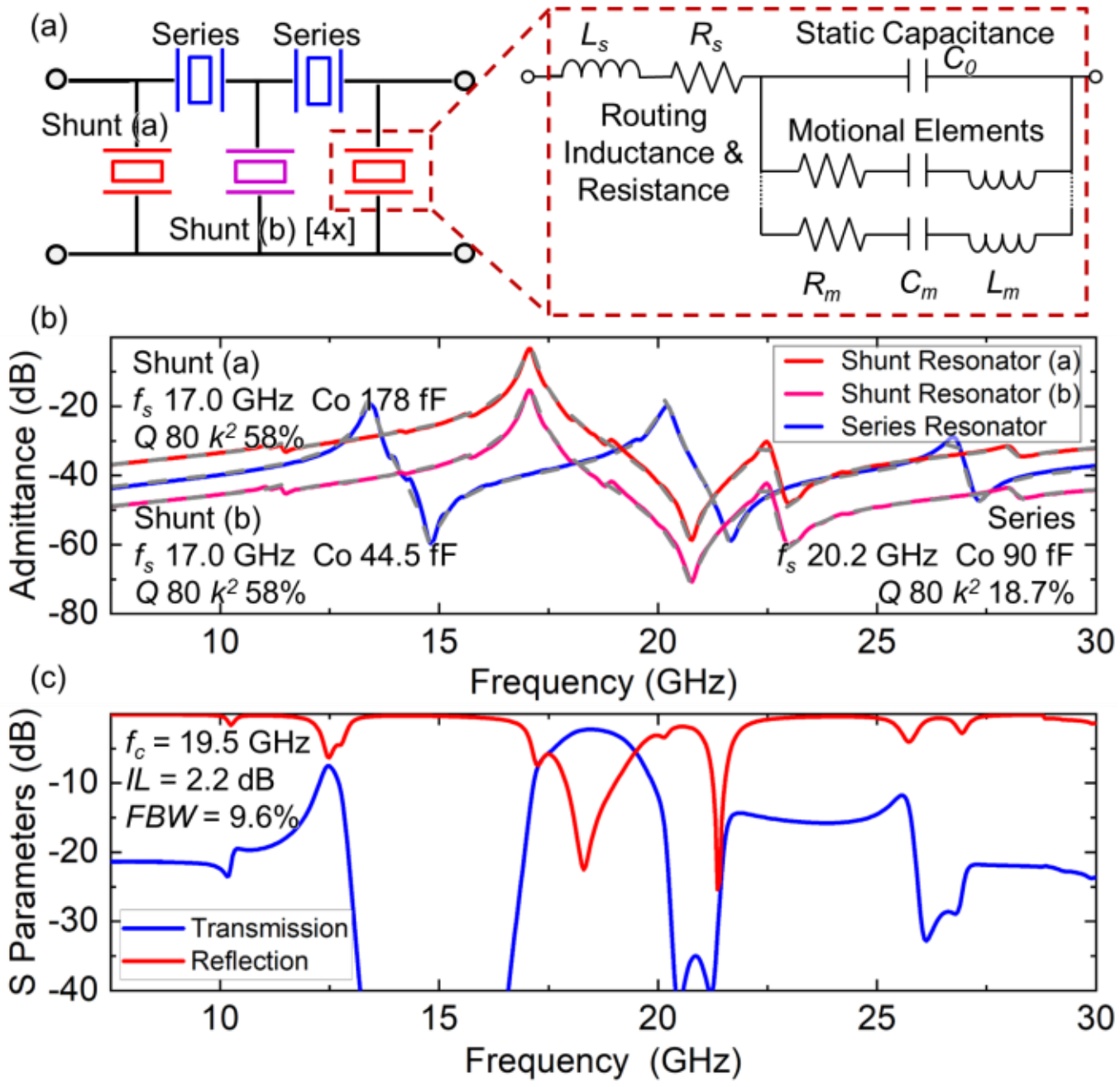

Fig. 2. (a) Circuit schematic of 5th-order ladder filter, and equivalent mBVD model at mmWave. (b) FEA simulated resonator response, and (c) synthesized filter transmission and reflection.

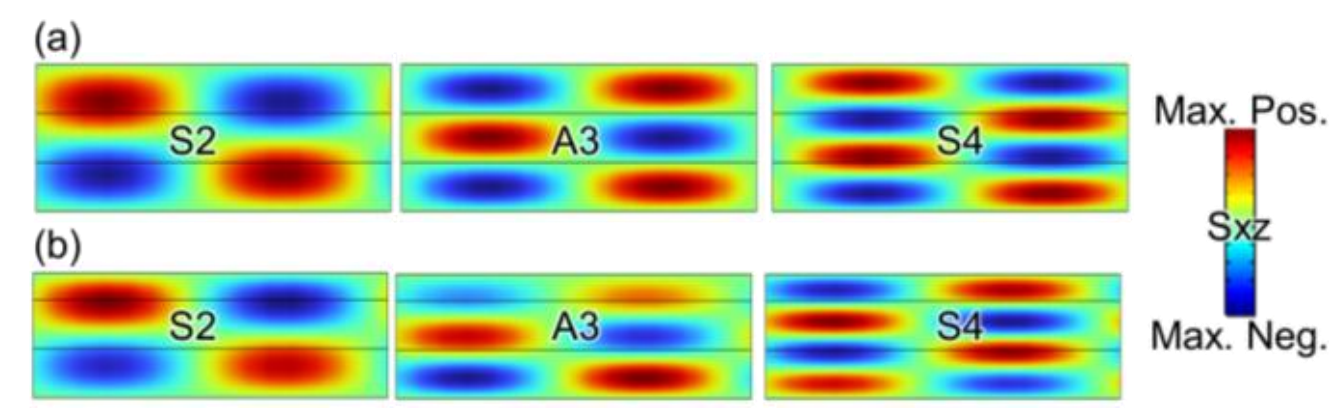

Fig. 3. FEA simulated S2, A3, and S4 stress profiles in trilayer P3F LN with (a) equal-thickness layers, and (b) a thinned top layer. The stress distributions for S2 and S4 modes lose symmetry in the thinned top-layer configuration (b), enhancing coupling and exciting these adjacent modes.

## II. Device Design and Fabrication

The filter is built on a 310 nm trilayer 128° Y-cut LN on sapphire carrier wafer, with a 1 µm-thick intermediate amorphous silicon sacrificial layer. In P3F LN, adjacent layers with opposite orientations support the third-order antisymmetric (A3) mode around 18 GHz (Fig. 1). More details on the stack are reported in [24].

A fifth-order ladder filter is designed [Fig 2 (a)], a Π-type ladder topology was selected in this design primarily due to simplified interconnect routing and grounding, minimizing layout complexity at mmWave frequencies. The frequency shifting between series and shunt resonators is achieved by local LN trimming of the top LN layer. More specifically, the series resonators have 260 nm LN, while the shunt resonators have 310 nm thick LN. The filter was fabricated following the procedure reported in [19], with specific details on ion milling provided in [25]. The dimensions of resonators are listed in Table II, which includes three types of resonators in the ladder topology. Ideally, in a tri-layer P3F LN with uniform thickness between layers, only A3 is predominantly excited, while second-order symmetric (S2) and fourth-order symmetric (S4) modes have minimal coupling, due to stress cancellation [Fig. 3(a)]. However, when the top layer of LN is trimmed for frequency shifting between the series and shunt resonators, S2 and S4 will obtain notable coupling [Fig. 3 (b)] [26]. Conventionally, P3F film-based filters avoid such effects by initially starting with a thicker top layer for the shunt resonator. In this work, we propose a novel approach to employ adjacent modes for achieving high close-in rejection in filter design.

Table II. Resonator Design Parameters

| Resonator | LN Thick. (nm) | Lateral Wave-Length (µm) | Electrode Width (nm) | Electrode Pair No. | Aperture (µm) |
|---|---|---|---|---|---|
| Series | 260 | 12 | 800 | 12 | 71 |
| Shunt (a) | 310 | 8 | 800 | 19.5 | 71 |
| Shunt (b) | 310 | 8 | 800 | 6.5 | 71 |

Table III. Simulated Resonator Parameters Near Passband

| Resonator | $C_0$ (fF) | Mode | $f_s$ (GHz) | $Q$ | $k^2$ (%) |
|---|---|---|---|---|---|
| Series | 90 | S2 | 13.4 | 80 | 27.8 |
| | | A3 | 20.2 | 80 | 18.7 |
| | | S4 | 26.8 | 80 | 6.54 |
| Shunt (a) | 178 | S2 | 11.5 | 80 | 2.4 |
| | | A3 | 17.1 | 80 | 58.7 |
| | | S4 | 22.5 | 80 | 5.7 |
| Shunt (b) | 44.5 | S2 | 11.5 | 80 | 2.4 |
| | | A3 | 17.1 | 80 | 58.7 |
| | | S4 | 22.5 | 80 | 5.7 |

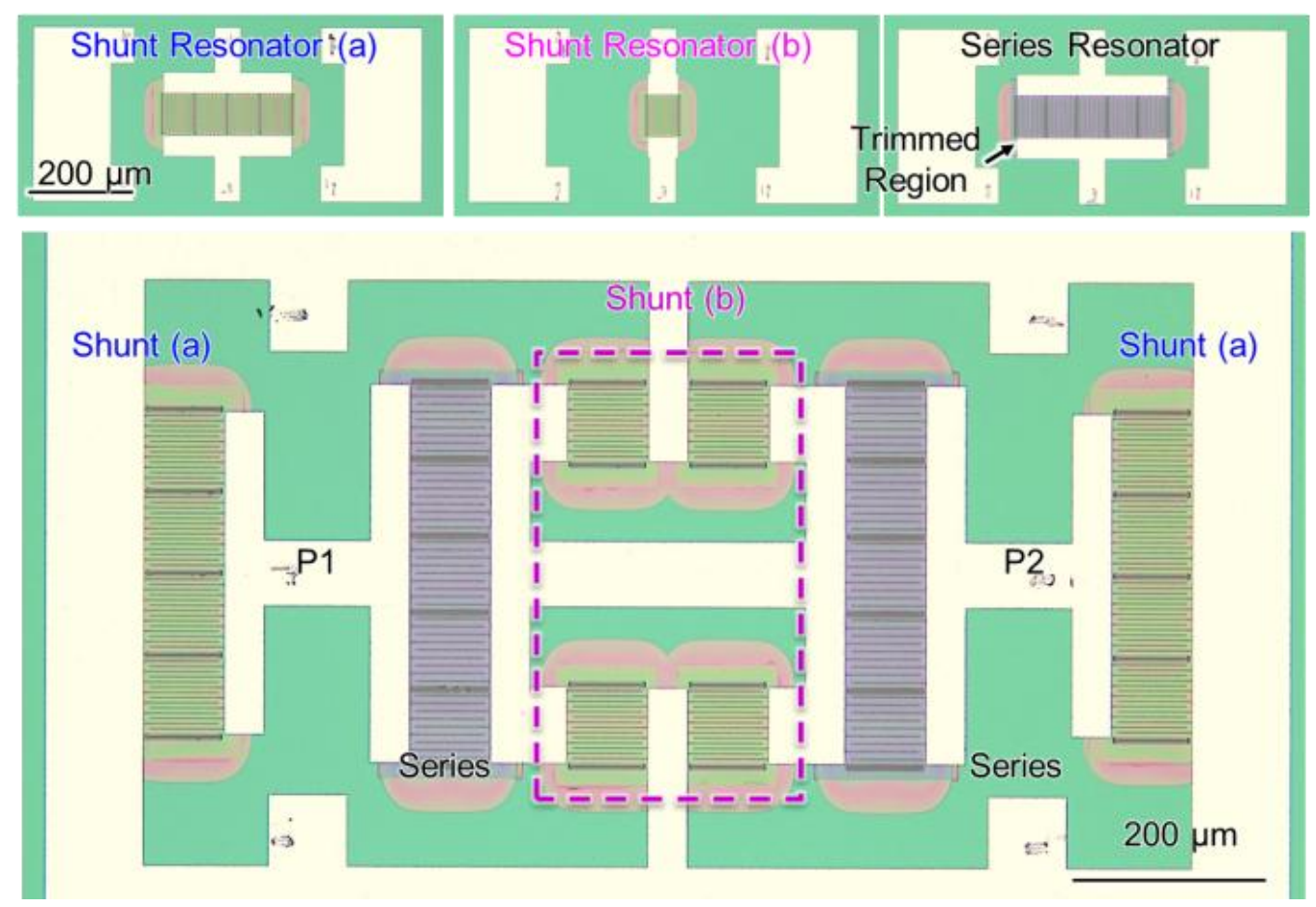

Fig. 4. Optical image of standalone resonator testbeds and filter results.

Resonator unit cells are initially simulated in COMSOL [Fig. 2(b)], with baseline parameters obtained from previously fabricated resonators [24]. The FEA model includes both surrounding air and substrate effects to account for EM feedthrough. The simulated Q was empirically selected based on our previously LN resonators at similar frequencies. Key extracted resonator parameters from these simulations are then imported into ADS, where the device dimensions are designed towards the desired bandpass and fitted with a multi-branch modified Butterworth-Van-Dyke (mBVD) model [Fig. 2(a)]. Device dimensions were selected to target the desired bandpass well below 30 GHz, a range where resonator self-resonance arising from parasitic inductances can become significant [19]. It is notable that despite the shunt resonator having a clean spectrum with pronounced S3, the series resonator shows obvious second and fourth order symmetric (S2 and S4) modes, intrinsically occurring in P3F resonators with thinned top LN. The resonator results are then fitted [dashed lines in Fig. 2 (b)], and the key parameters are extracted in Table III, highlighting the coexistence of S2, A3, and S4 modes.

The simulated S parameters of resonators are then imported into Keysight Advanced Design System for filter simulation. The designs are optimized for a minimum IL of 2.2 dB and FBW of 9.6% with 50Ω terminations [Fig. 2(c)]. Remarkably, the design features close-in rejection bands below the passband, with over 40 dB rejection over a 2.9 GHz bandwidth. The high rejection is accomplished by the adjacent S2 and S4 modes, flipping the phase of resonators near the main passband, and creating band-stop performance in the close-in range.

## III. Measurement and Discussion

The filter is fabricated following the procedure reported in [19]. The electrodes are defined via electron-beam lithography (EBL) and consist of 300 nm-thick aluminum (Al), with widths of 800 nm to achieve spurious mode suppression [27], and with the buslines thickened by an additional 300 nm of Al for reduced routing resistance ($R_s$). The optical images of the fabricated resonator testbeds and filter are shown in Fig. 4. The devices are then measured in air with a network analyzer. A GGB CS-5 calibration substrate was used to define the measurement reference plane at the probe tips. The admittance curves of resonators are plotted in Fig. 5 (a). The key parameters are extracted and plotted in Table IV, showing good matching to the simulation, with a small difference caused by the minor film thickness variation between layers. Note that the reported $Q_s$ values are significantly affected by the electrical loading from $R_s$, whereas the $Q_p$ is less impacted, thus more suitable for comparison to simulated results. Additionally, resonators with larger $C_0$ and $k^2$ exhibit lower perceived $Q_s$ due to increased susceptibility to EM loading effects.

The filter response is plotted in Fig. 5 (b), exhibiting an IL of 2.2 dB and a 3-dB FBW of 8.5% at 19.3 GHz with 50 Ω impedance. The zoom-in passband result [Fig. 5 (c)] shows 49.9 dB close-in rejection, measured at twice the 3-dB FBW offset from the center frequency, a representative spacing adopted as a potential requirement for diplexer applications. The higher close-in rejection below the passband (14.5–17 GHz) results from multiple transmission zeros, specifically the series resonance ($f_s$) from the A3 mode contribution of 3 shunt resonators and the S2 mode parallel resonance ($f_p$) of the series resonators. In contrast, above the passband, mainly the two series resonators contribute through the A3 mode $f_p$ resonance, thereby limiting the rejection. A comparison to the state-of-the-art (Table I) using this criterion shows the frequency scaling of acoustic filters comparable to prior works, and with substantial improvement in the close-in rejection bands. A drawback of the proposed approach is the reduced rejection performance at frequencies far from the passband. Future work will focus on enhancing overall rejection by optimizing the P3F film stack to exploit additional adjacent modes, such as the S4 mode above the passband. We also plan to investigate advanced filter topologies, including hybrid EM–acoustic designs, to improve out-of-band rejection beyond the immediate vicinity of the passband. [28]. Moreover, future studies will examine electrode aging and oxidation effects on filter performance, as well as suitable fabrication strategies to mitigate potential impacts.

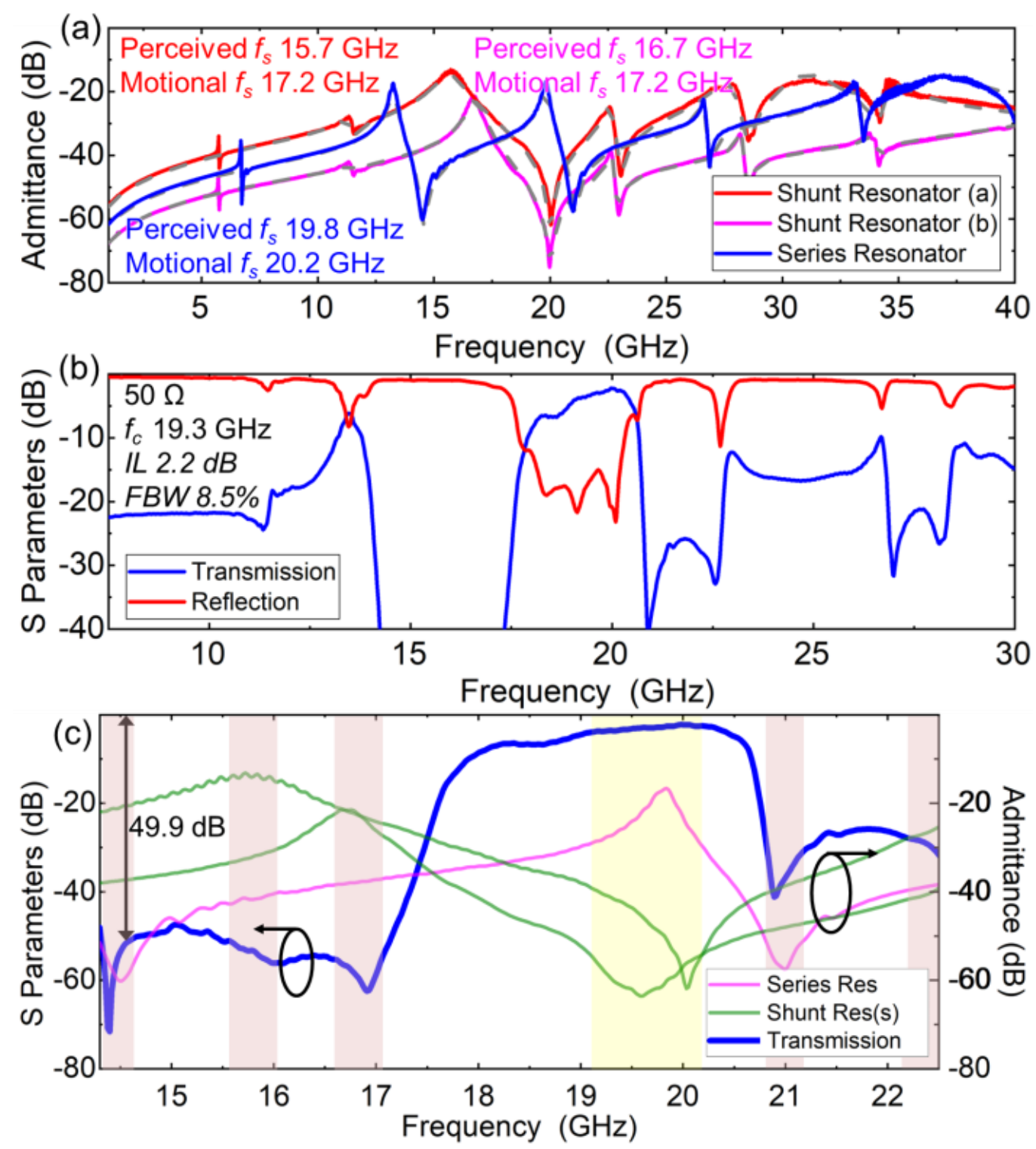

Fig. 5. (a) Measured resonator response, (b) filter transmission and reflection. (c) Zoom-in S parameter of the passband, highlighting the high close-in rejection enabled by P3F LN.

Table IV. Measured Resonator Parameters Near Passband

| Resonator | $C_0$ (fF) | $L_s$ (nH) | $R_s$ (Ω) | Mode | $f_s$ (GHz) | $Q_s$ | $Q_p$ | $k^2$ (%) |
|---|---|---|---|---|---|---|---|---|
| Series | 96 | 0.21 | 6 | S2 | 13.5 | 53.8 | 72 | 24.2 |
| | | | | A3 | 20.2 | 87.7 | 105 | 10 |
| | | | | S4 | 26.7 | 136 | 285 | 1.2 |
| Shunt (a) | 174 | 0.19 | 5 | S2 | 11.4 | 12.9 | 11.3 | 2.8 |
| | | | | A3 | 17.2 | 18.9 | 225 | 46 |
| | | | | S4 | 22.7 | 46.6 | 138 | 1.9 |
| Shunt (b) | 48 | 0.24 | 9.4 | S2 | 11.5 | 10.2 | 13.4 | 0.8 |
| | | | | A3 | 17.2 | 36.3 | 266 | 45 |
| | | | | S4 | 22.7 | 78.5 | 140 | 1.2 |

## IV. Conclusion

This work demonstrates the first tri-layer P3F LN acoustic filter operating at 19.3 GHz, achieving a low IL of 2.2 dB, a 3-dB FBW of 8.5%, and a state-of-the-art close-in rejection of 49.9 dB. By leveraging the intrinsic multi-mode behavior of the tri-layer P3F LN structure and introducing thickness trimming to precisely control the resonator spectra, the filter design achieves strong band-stop behavior adjacent to the passband, without requiring additional resonators. This novel approach offers a promising path toward compact, high-rejection diplexer and multiplexer solutions for future FR3 band filters.